\documentclass[12pt]{article}

\input epsf

\def\Im{\mbox{Im}}
\def\Re{\mbox{Re}}

\title{Pair production of fundamental unstable particles in modified
perturbation theory in NNLO}
\author{M. L. Nekrasov \\
{\small \it Institute for High Energy Physics, 142280 Protvino,
Russia}}

\date{}
\begin{document}

\maketitle

\begin{abstract}
We consider pair production and decay of fundamental unstable
particles in the framework of a modified perturbation theory (MPT)
treating resonant contributions in the sense of distributions. The
cross-section of the process is calculated within the NNLO of the
MPT in a model that admits exact solution. Universal
massless-particles contributions are taken into consideration. A
comparison of the outcomes with the exact solution demonstrates
excellent convergence of the MPT series at the energies near and
above the maximum of the cross-section.

\end{abstract}

\section{Introduction}\label{int}

A description of the processes of productions and decays of
fundamental unstable particles to satisfy the up-to-date
requirements must provide, on the one hand, gauge cancellations
and unitarity and, on the other hand, enough high accuracy of
calculation of resonant contributions of unstable particles.
Unfortunately, in the framework of conventional perturbation
theory (PT) a simultaneous fulfilling of these requirements is
obstructed by divergences caused by resonant contributions. For
this reason in the propagators of unstable particles the Dyson
resummation is usually applied, which shifts the resonant
singularities out of the region of physical momenta. However, a
resummation mixes the PT orders, which generally leads to
violation of the gauge cancellations. So simultaneously with using
the Dyson resummation an application of additional tricks is
required.

Among various approaches that include such tricks, the most known
one is based on the Laurent expansion of the amplitude around the
complex poles of the resonant propagators. Each term of this
expansion is considered expanded in the framework of the
conventional PT, as well, but a certain portion of the self-energy
is not involved in the latter expansion as having been absorbed by
the shift of the point of singularity (the remnant of the Dyson
resummation). The gauge cancellations are completely maintained in
this approach. However, the precision of the description vastly
falls at the increasing of a distance from the resonant region,
and an uncertainty arises at calculating the residues in the
complex-poles. Nevertheless, in the vicinity of the resonant
region the pole expansion in many cases is suitable for
applications. In particular, at LEP2 the loop corrections to the
$W$-pair production were calculated in the double pole
approximation (DPA) \cite{LEP2a,LEP2b}, the leading approximation
in the pole expansion. Unfortunately, at international linear
collider (ILC) \cite{ILC} the accuracy of DPA is no longer
sufficient \cite{CMS1}, and the higher-order corrections in the
pole expansion unlikely can save the situation. Therefore the
pinch-technique method and the method based on the
background-field formalism move forward to foreground, which, in
principle, can provide the necessary precision (see \cite{pinch}
and \cite{BFM}, and the references therein). However, the
consecutive application of the mentioned methods implies a
calculation of a huge volume of additional contributions that
formally appear outside the limits of required precision, which is
impractical \cite{Ditt}. So at present hopes are pinning on the
approach of ``complex-mass scheme'' (CMS), which avoids mentioned
difficulties \cite{CMS1,CMS2}. Nevertheless, in the CMS another
problem related to the unitarity arises. The point is that the CMS
uses the complex-valued renormalized masses for unstable particles
and this requires an introduction of the complex-valued
counterterms, which violates unitarity. For this reason the CMS
cannot be considered as a rigorous procedure \cite{CMS2}. The
problem becomes especially topical at calculating the
contributions in the next-to-next-to-leading order (NNLO). Thus to
make the calculations up to the NNLO alternative approaches are
required.

A promising candidate for this role is a modified perturbation
theory (MPT), first proposed in \cite{Tkach} and then elaborated
in \cite{EPJC} and \cite{MPT}. For determining the resonant
contributions the distribution theory is applied in this approach
instead of the Dyson resummation in whatever form. In essence, the
MPT implies a systematic expansion in powers of the coupling
constant directly of the probability instead of the amplitude.
This mode allows one to impart the sense of distributions to the
propagators squared of unstable particles, and on this basis to
asymptotically expand the propagators squared without the
appearance of the divergences in the cross-section. Since the
object to be expanded (the cross-section) is gauge invariant and
the expansion is made in powers of the coupling constant, the
result of the expansion must automatically be gauge-invariant.
This implies that the gauge cancellations in the MPT must be
automatically maintained. Of course, this should be so if the MPT
exists, i.e. if it is a well-determined method. In the case of
pair production of unstable particles this property was proved and
an algorithm of the calculation of each order of the MPT expansion
was elaborated \cite{MPT}.

The aim of the given paper is to perform numerical analysis of the
convergence properties of the MPT series in the case of pair
production of unstable particles. At once we should notice that in
the qualitative sense the outcomes should weakly depend on the
model under consideration because the choice of a model implies
mainly a definition of the test function in the presence of which
the relevant distributions (the propagators squared) are
MPT-expanded. So it is reasonable to carry out the examination in
the framework of a model possessing an exact solution. As such a
model, we consider the improved Born approximation for the process
$e^{+} e^{-} \to \gamma,Z \to t\bar t \to W^{+} b\:W^{-}\bar b$.
For simplicity we consider $W$ bosons and $b$ quarks to be stable
particles, with $W$ being massive and $b$ being massless. At the
same time we consider realistic corrections to the width of the
top quark. This should allow us to get an information about the
rapidity of convergence in the realistic case. A similar model has
actually been considered in \cite{N-tt} by examining the MPT
within the next-to-leading order (NLO). However, the contribution
from the soft massless particles to the process of production of
unstable particles have been omitted in that work. This was a
serious omission because the mentioned contributions include
Coulomb singularities \cite{Coulomb1}-\cite{Fadin1} appreciably
affecting the cross-section. In this paper we improve the
calculations of \cite{N-tt} (in particular eliminate some bug in
the calculations), and carry out numerical calculations further up
to the NNLO with taking into consideration universal Coulomb
singular contributions.

In the next section, we present the basic information about the
MPT and detail the model in the framework of which we carry out
computations. In Sect.~\ref{num} we present outcomes. In
Sect.~\ref{Conclusion} we discuss the results.

\section{MPT and a model for its examination}\label{mod}

The observable cross-section of production and decay of unstable
particles, for example in $e^+ e^-$ annihilation, has the form of
a convolution of the hard-scattering cross-section with the flux
function \cite{LEP2a},
\begin{equation}\label{not1}
\sigma (s) = \int\limits_{s_{\mbox{\tiny min}}}^s \frac{\mbox{d}
s'}{s} \: \phi(s'/s;s) \> \hat\sigma(s')\,.
\end{equation}
Here $s$ is the energy squared in the center-of-mass system,
$\hat\sigma$ is the hard-scattering cross-sections, $\phi$ is the
flux function describing contributions of nonregistered photons
emitted in the initial state. The $s'/s$ characterizes a fraction
of the energy expended on the production of unstable particles.
For our purposes it is sufficient to take $\phi$ in the
leading-log approximation. So we put
\begin{equation}\label{not2}
\phi (z;s) = \beta_e (1-z)^{(\beta_e -1)}
             - \frac{1}{2} \beta_e (1+z), \qquad
\beta_e = \frac{2 \alpha }{\pi}\left(\ln \frac{s}{m_e^2}
-1\right).
\end{equation}

In the case of pair production of unstable particles the
double-resonant contributions are most crucial. Bearing this in
mind we write down the hard-scattering cross-section in the form
\begin{equation}\label{not3}
\hat\sigma (s) \; = \!\!\!\!
 \int\limits_{\quad {\displaystyle\mbox{\scriptsize $s$}}_
                    { 1 \mbox{\tiny min} } \atop
              \quad {\displaystyle\mbox{\scriptsize $s$}}_
                    { 2 \mbox{\tiny min} }}
            ^{\!\!\infty}   \!\!\!\!\!\!\!\!
 \int\limits^{\infty} \mbox{d} s_1 \, \mbox{d} s_2 \;\;
 \hat\sigma(s\,;s_1,s_2) \left(1\!+\!\delta_{c}\right).
\end{equation}
In this formula $\hat\sigma(s\,;s_1,s_2)$ is an exclusive
cross-section, $\delta_{c}$ stands for soft massless-particles
contributions, $s_1$ and $s_2$ are virtualities of unstable
particles. In the case of the process $e^+e^- \to \gamma,Z \to
t\bar t \to W^{+}b\:W^{-}\bar b$ with massive $W$ and massless
$b$, we have $s_{1\,\mbox{\scriptsize min}} =
s_{2\,\mbox{\scriptsize min}} = M_{W}^2$, and
$s_{\mbox{\scriptsize min}}=4M_W^2$. In $\hat\sigma(s\,;s_1,s_2)$
we extract kinematic and Breit-Wigner (BW) factors,
\begin{equation}\label{not4}
\hat\sigma(s\,;s_1,s_2) = \frac{1}{s^2} \,
  \theta(\sqrt{s}-\!\sqrt{s_1}-\!\sqrt{s_2}\,)
  \sqrt{\lambda(s,s_{1},s_{2})}\;\Phi(s;s_1,s_2)
  \> \rho(s_{1}) \> \rho(s_{2})\,.\quad
\vspace*{0.4\baselineskip}
\end{equation}
Here $\lambda (s,s_{1},s_{2}) = [s \!-\!(\sqrt{s_1} \!+\!
\sqrt{s_2}\,)^2] [s \!-\!(\sqrt{s_1} \!- \! \sqrt{s_2}\,)^2]$ is
the kinematic function, $\rho(s_{1})$ and $\rho(s_{2})$ are BW
factors. Function $\Phi(s;s_1,s_2)$ is the rest of the amplitude
squared. Below we consider $\Phi$ in the Born approximation, and
thus only the BW factors are subject to the MPT expansion. In the
general case, we define the BW factors as
\begin{equation}\label{not5}
\rho (s) = \frac{M \Gamma_{0}}{\pi}\times |\Delta(s)|^2\,.
\end{equation}
Here $M$ is the renormalized mass of the top quark, $\Gamma_{0}$
is its Born width, $\Delta(s)$ is a scalar part of the
Dyson-resummed propagator (the spin factor is referred to $\Phi$),
\begin{equation}\label{not6}
\Delta^{-1}(s) =
 s - M^2 + \Re \Sigma(s) + \mbox{i}\:\Im \Sigma(s)\,,
\end{equation}
$\Re\Sigma(s)$ and $\Im\Sigma(s)$ are the real and imaginary parts
of the renormalized self-energy.

In the case of a smooth weight, an isolated BW factor may be
represented in the form of an asymptotic expansion in the sense of
distributions in powers of the coupling constant (generating thus
the MPT expansion of isolated BW factor). Up to and including the
NNLO this expansion looks as follows \cite{Tkach}:
\begin{eqnarray}\label{not7}
&{\displaystyle \rho(s) \;=\; \delta(s\!-\!M^2) \,+\,
 \frac{M \Gamma_{0}}{\pi} \, PV \! \left\{\frac{1}{(s-M^2)^2} -
 \frac{2\alpha\,\Re\Sigma_1(s)}{(s\!-\!M^2)^3}\right\}}&
 \\
&\displaystyle +\;\sum\limits_{n\,=\,0}^2 c_{n}(\alpha)\,
 \frac{(-)^{n}}{n!}\,\delta^{(n)}(s\!-\!M^2) + O(\alpha^3)\,.&\nonumber
\end{eqnarray}
Here $\alpha$ is the coupling constant, $\delta(\cdots)$ is the
$\delta$-function, $\delta^{(n)}$ is its $n$th derivative, $PV$
means the principal-value prescription. The leading term in
(\ref{not7}) defines the narrow-width approximation. The
contributions in the curly brackets appear as a result of the
naive expansion of the propagator squared; $PV$ makes the poles in
this expansion integrable. The contributions under the sum-sign
correct the contributions of the $PV$ poles (in the singular
point) so that the expansion becomes asymptotic. The coefficients
$c_{n}(\alpha)$ are polynomials in $\alpha$, determined by the
self-energy contributions of the unstable particle. In an
arbitrary UV-renormalization scheme the completely explicit
expressions for $c_{n}(\alpha)$ may be found in \cite{EPJC}. In
the case of the on-mass-shell (OMS) type scheme, they are found in
\cite{MPT}. In the latter case the coefficients $c_n$ within the
NNLO are determined by $I_1$, $I_2$, $I_3$, $I'_1$, $I''_1$, where
$I_n=\Im\Sigma_n(M^2)$, $I'_n=\Im\Sigma'_n(M^2)$,
$I''_n=\Im\Sigma''_n(M^2)$, and by $R_2$, $R'_2$, where
$R_n=\Re\Sigma_n(M^2)$, $R'_n=\Re\Sigma'_n(M^2)$. Here $\Sigma_n$
is the $n$-loop self-energy defined in accordance with relation
$\Sigma = \alpha \Sigma_1 + \alpha^2 \Sigma_2 + \cdots$.

Unfortunately, the weight in our case is not smooth because of the
kinematic factor in formula (\ref{not4}). A solution to this
problem is found on the basis of analytic regularization via the
substitution $[\lambda (s,s_{1},s_{2})]^{1/2} \to [\lambda
(s,s_{1},s_{2})]^{\nu}$. Furthermore, the weight
$\Phi\,(1+\delta_{c})$ may be expanded in powers of $s_1$ and
$s_2$ around $s_1=M^2$ and $s_2=M^2$. Then, it becomes possible to
analytically calculate singular integrals irrespective of details
of the definition of the weight. After calculating singular
integrals and removing the regularization the outcomes remain
finite and the expansion remains asymptotic \cite{MPT}. In
principle, this salvages the applicability of the approach, and
the problem is reduced to numerical calculations only.

Now we turn to the definition of the model in the framework of
which we will carry out calculations. At first we notice that
within the NNLO the MPT expansion of the general BW factor based
on propagator (\ref{not6}) coincides with that based on the
following ``minimal'' propagator:
\begin{eqnarray}\label{not8}
\Delta^{-1}_{NNLO}(s) &=& s - M^2 + \alpha \, \Re \Sigma_1(s) +
\mbox{i}\,\alpha \left[I_1 + (s \!-\! M^2)\,I'_1
+ \frac{1}{2}(s \!-\! M^2)^2 I''_1 \right]\nonumber\\
& + & \alpha^2 \left[ R_2 + \mbox{i}\, I_2 + (s \!-\! M^2)\,R'_2
\right] + \mbox{i}\,\alpha^3 I_3\,.
\end{eqnarray}
In fact, after the MPT expansion any contribution not included in
(\ref{not8}) appears outside the NNLO. However, under the
consideration in the conventional-function sense, the $R_3$ and
$I'_2$ in the region $s-M^2 \sim O(\alpha)$ may be assigned to the
NNLO, as well. So we start from the following modeling propagator:
\begin{eqnarray}\label{not9}
\Delta^{-1}_{NNLO}(s) & = & s \, - \, M^2 \, + \, \alpha \, \Re
\Sigma_1(s) \, + \, \mbox{i}\,\alpha \, \Im \Sigma_1(s)
\nonumber\\[0.5\baselineskip]
&+& \alpha^2 \left[R_2 + \mbox{i} \, I_2 + (s-M^2) (R'_2 +
\mbox{i} \, I'_2)\, \right] \: + \: \alpha^3 \left(R_3 + \mbox{i}
\, I_3 \right).
\end{eqnarray}
Propagator (\ref{not9}) ensures the NNLO precision from the point
of view of both the MPT and conventional functions. For uniformity
we consider $\Im\Sigma_1$ off-shell as well as in the case of
$\Re\Sigma_1$.

Now let us direct our attention to the definition of the two- and
three-loop contributions to propagator (\ref{not9}). Actually they
may be determined by using the only fact that they are the
on-shell contributions. Specifically, the real parts may be
determined by basing on the UV-renormalization conditions. But one
should remember that in the unstable-particles case the OMS scheme
may be determined in different fashions. In particular, the
conventional OMS scheme is determined by the conditions $R_n=0$
and $R'_n=0$ \cite{OMS}. However, it is inconvenient for the
calculations in the higher-orders, because the renormalized mass
$M$ in this scheme beginning with the two-loops is different from
the observable mass, and beginning with the two-loops generally is
gauge-dependent \cite{Sirlin1}. The problem is eliminated at
considering the first renormalization condition in the form $M^2 =
\Re \, s_p\,$, where $s_p$ is the pole of the propagator,
$\Delta^{-1}(s_p) = 0$, which means the equating of the
renormalized mass to the observable mass. The second
renormalization condition may be determined by equating the
imaginary part of the on-shell self-energy to the imaginary part
of $s_p\,$. As a result the equality $s_p = M^2 - \mbox{i} \, I$
is established by means of the UV-renormalization conditions,
where $I =\Im \Sigma(M^2)$. So both the renormalized mass and the
imaginary part of the on-shell self-energy become
gauge-independent. This scheme of the UV renormalization was
called the $\overline{\mbox{OMS}}$ scheme in \cite{OMS-bar} and
the ``pole scheme'' in \cite{Sirlin2}. In this scheme the $R_2$,
$R'_2$ and $R_3$ are determined as
\begin{equation}\label{not10}
 R_2 = -I_1 I'_1 \,,
\qquad R'_2= - I_1 I''_1 / 2 \,, \qquad
 R_3 = - I_2 I'_1 - I_1 I'_2 + I^2_1 R''_1 / 2\,.
\end{equation}

The imaginary contributions to the on-shell self-energy are
determined, in effect, by the unitarity condition. For $I_1$ and
$I_2$ the appropriate relations are
\begin{equation}\label{not11}
\alpha I_1 = M\Gamma_0\,, \qquad\quad \alpha^2 I_2 = M \alpha
\Gamma_1 \, .
\end{equation}
Here $\Gamma_0$ and $\alpha\Gamma_1$ are the Born and the one-loop
contributions to the width. The $I_3$, in the general case, is
nontrivially related with the two-loop contribution
$\alpha^2\Gamma_2$. In the $\overline{\mbox{OMS}}$ scheme this
relation is \cite{OMS-bar}
\begin{equation}\label{not12}
\alpha^3 I_3 = M \alpha^2 \Gamma_2 + \Gamma_0^3/(8M) \,.
\end{equation}

Unfortunately, the derivatives of $\Im \Sigma$ cannot be
determined by similar means. However the $I'_2$, which we need, is
a facultative quantity from the point of view of the MPT expansion
(see above). So we may determine $I'_2$ by using the approximate
relationships $\alpha^2 \Im\Sigma_2(s) = \sqrt{s}\,\alpha
\Gamma_1(s)$, $\Gamma_1(s) = \Gamma_1 \times
\Gamma_0(s)/\Gamma_0$, $\Gamma_0(s) = \alpha
\Im\Sigma_1(s)/\!\sqrt{s}$. This yields
\begin{equation}\label{not13}
\alpha^2 I'_2 = \frac{\alpha \Gamma_1}{\Gamma_0} \;\, \alpha I'_1
\,.
\end{equation}

Now we determine the one-loop self-energy $\Sigma_1(s)$. In the
framework of the model, we determine it with contributions of the
$W$ boson and $b$ quark only. In this way we avoid the IR
divergences generally arising at determining $\Re\Sigma_1(s)$.
Standard calculation in t'Hooft-Feynman gauge\footnote{The gauge
independence should be restored after including the higher-order
corrections in $\Phi$, and after including the single- and
non-resonant contributions in the cross-section \cite{MPT}.} gives
\begin{equation}\label{not14}
\alpha \, \Sigma_1(s) = A(s) - \Re A(M^2) - (s-M^2) \Re A'(M^2)\,,
\end{equation}
\begin{equation}\label{not15}
 A (s) = - \frac{G_F M_W^2}{4\sqrt{2}\,\pi^2}\,s\left[
 \left(2+\frac{M^2}{M_W^2}\right) B_1(s;0,M_{W}) + 1\right].
\end{equation}
Here $B_1(s;m_1,m_2)$ is the Passarino-Veltman function \cite{BP}.

Thus, we have determined all contributions to the propagator
(\ref{not9}) and thereby the BW factors in formula (\ref{not4}).
Further, by virtue of (\ref{not7}) we can determine the MPT
expansion of the BW factors. The coefficients $c_n$ on account of
(\ref{not10})--(\ref{not12}), (\ref{not14}), (\ref{not15}) and
\cite{MPT} are as follows:
\begin{eqnarray}\label{not16}
c_0 & = & - \, \alpha \, \frac{\Gamma_1}{\Gamma_0} + \alpha^2
\left[\frac{\Gamma_1^2}{\Gamma_0^2} -
\frac{\Gamma_2}{\Gamma_0}\right] - \frac{\Gamma_0^2}{8 M^2} -
\left(\frac{\Gamma_0}{M} \, \frac{M^2+M_W^2}{M^2-M_W^2}\right)^2,
\nonumber \\[0.5\baselineskip]
c_1 &=& 0, \qquad\qquad c_2 \;\;=\; -\, M^2 \Gamma_0^2 \,.
\end{eqnarray}
Recall that each $\Gamma_n$ includes an additional factor
$\alpha$, which is conditioned by the vertex origin of the width.

To complete definition of the model, we must determine also the
factor $(1+\delta_{c})$ in formula (\ref{not3}). Let us remember
that we have ignored in the self-energy all massless-particles
contributions that lead to IR divergences. For this reason we have
to ignore all other soft-massless-particles contributions whose
IR-divergent contributions are to be cancelled in the
cross-section. So, there should remain only the Coulomb singular
contributions that are not cancelled. Recall that they have the
meaning of universal corrections arising due to exchanges by soft
massless particles (photons, gluons) between outgoing massive
particles in the limit of small relative velocities. In the case
of strong-interacting top quarks, it is reasonable to ignore the
exchanges by photons and to take into account only gluon
exchanges. We also restrict our consideration to the one-gluon
approximation. Then, with taking into account the off-shell and
finite-width effects, we have \cite{Bardin,Fadin1}
\begin{equation}\label{not17}
{\delta_c} = \frac{\kappa \, \alpha_s \pi}{2\beta}\left[ 1 -
\frac{2}{\pi}\,\arctan \! \left( \frac{|\beta_{M}|^2 -
\beta^2}{2\beta \, \Im\beta_{M}} \right)\right] .
\end{equation}
Here $\kappa = 4/3$ is the group factor, $\alpha_s$ is the strong
coupling constant, $\beta = s^{-1}\sqrt{\lambda (s,s_{1},s_{2})}$
is the velocity of the unstable particles in the c.m.f.,
$\beta_{M} = \sqrt{1-4 (M^2\!- \mbox{i} M \Gamma)/s}$. Further we
put $\Gamma = \Gamma_0$ in the latter formula. The energy-scale
dependence in $\alpha_s$, we take into consideration as described
in \cite{Fadin2}.

So, now the model is completely determined. The cross-section in
the model may be straightforwardly calculated. We call the result,
the ``exact'' solution. Simultaneously we can calculate the MPT
expansion of the cross-section and compare the outcome with the
``exact'' result. Ultimately the expansion should have the form
\begin{equation}\label{not18}
\sigma(s) = \sigma_{0}(s) \,+\, \alpha\,\sigma_{1}(s) \,+\,
\alpha^2 \sigma_{2}(s) \,+\, \cdots \,.
\end{equation}
Here $\sigma_{0}$ means the cross-section in the LO approximation,
$\alpha\,\sigma_{1}$ and $\alpha^2 \sigma_{2}$ mean the NLO and
NNLO corrections, respectively. So, the $\sigma_{01} = \sigma_{0}
+ \alpha\,\sigma_{1}$ and $\sigma_{012} = \sigma_{0} +
\alpha\,\sigma_{1} + \alpha^2 \sigma_{2}$ determine the NLO and
NNLO approximations. Similarly we denote the contributions to the
hard-scattering cross-section $\hat\sigma(s)$.

\section{Results of numerical calculations}\label{num}

Parameters of the model we determine as follows: $M = 175 \:
\mbox{GeV}$, $M_W = 80.4 \: \mbox{GeV}$, and we use the following
previously calculated input-data for the width \cite{Top}:
\begin{eqnarray}\label{not19}
 \Gamma_0 &=& 1.56 \; \mbox{GeV} \,, \nonumber \\
 \Gamma_0 + \alpha \Gamma_1 &=& 1.45 \; \mbox{GeV} \,, \nonumber \\
 \Gamma_0 + \alpha \Gamma_1 + \alpha^2 \Gamma_2 &=&
 1.42 \; \mbox{GeV} \,.
\end{eqnarray}
The $\Gamma = \Gamma_0 + \alpha \Gamma_1 + \alpha^2 \Gamma_2$ we
consider as the total width of the top quark. From (\ref{not19}),
we get $\alpha \Gamma_1=-0.11$~GeV, and $\alpha^2
\Gamma_2=-0.03$~GeV. The $\alpha$ in $\Phi(s;s_1,s_2)$ and
$\phi(z;s)$, we set equal $1/137$. All calculations are carried
out on the basis of rather general FORTRAN code with double
precision written in accordance with the formulas and instructions
described in \cite{MPT}.

\begin{figure}[t]
\hbox{ 
       \epsfxsize=420pt \epsfbox{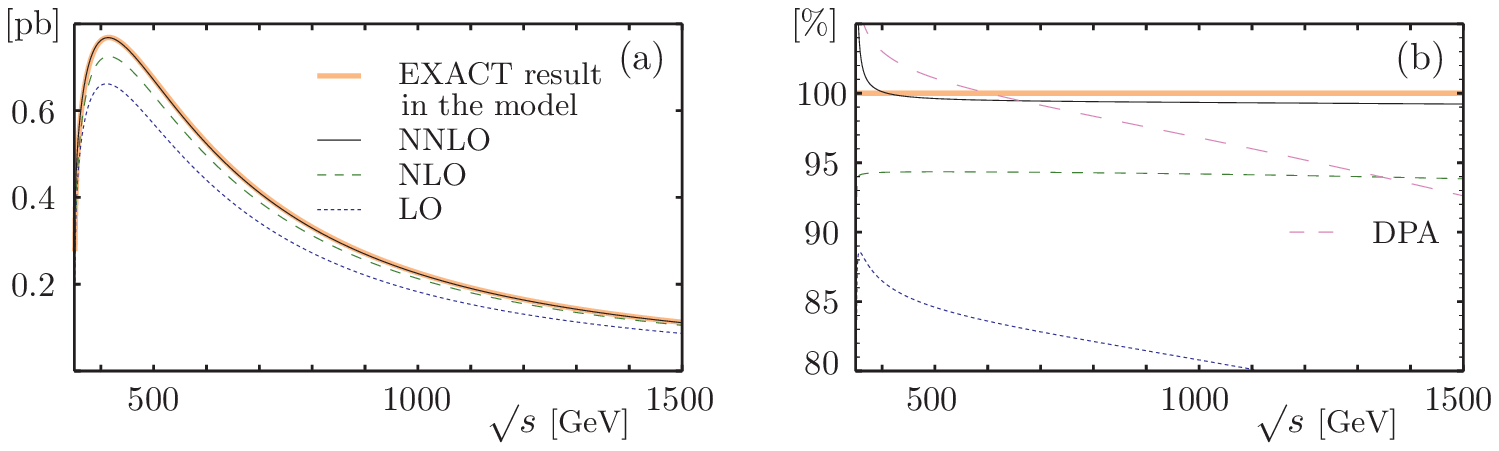}}
\caption{\small Total cross-section $\sigma(s)$. The exact result
in the model, we show by thick~curve. Dotted, short-dashed, and
continuous thin curves mean the LO, NLO, and NNLO approximations
in the MPT, respectively. The results are presented in pb (a) and
in percents to the exact result (b).  In panel (b), we show by the
long-dashed curve the result in the DPA.\label{Fig1}}
\end{figure}
\begin{figure}[t]
\hbox{ 
       \epsfxsize=420pt \epsfbox{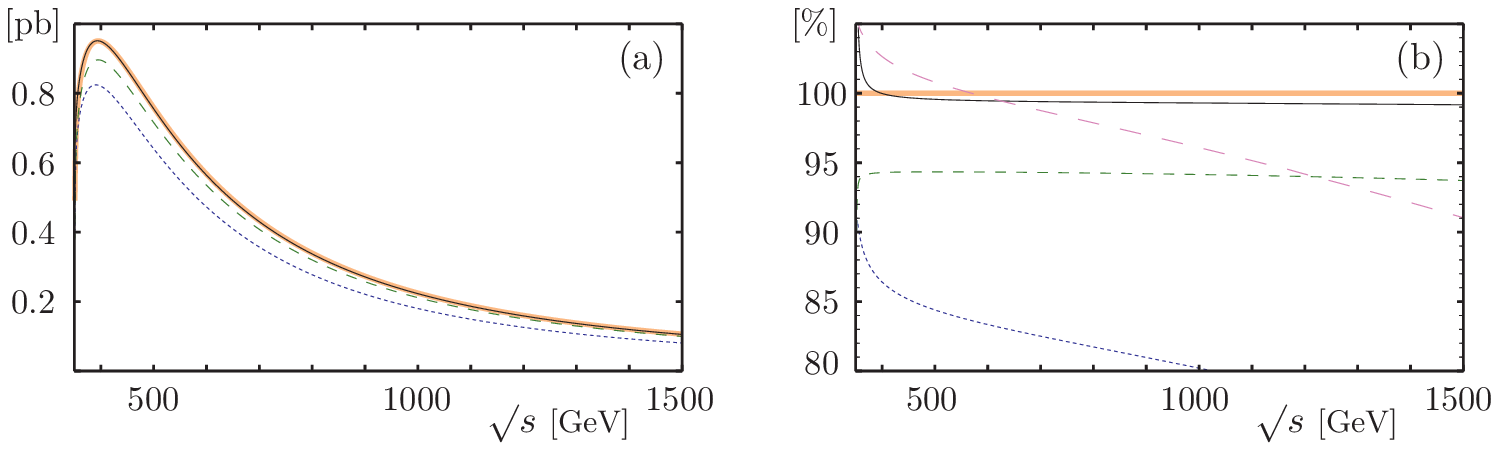}}
\caption{\small Hard-scattering cross-section $\hat\sigma(s)$. The
notation is the same that in Fig.~\ref{Fig1}.\protect\label{Fig2}}
\end{figure}

\begin{figure}[t]
\hbox{ \hspace*{80pt}
       \epsfxsize=260pt \epsfbox{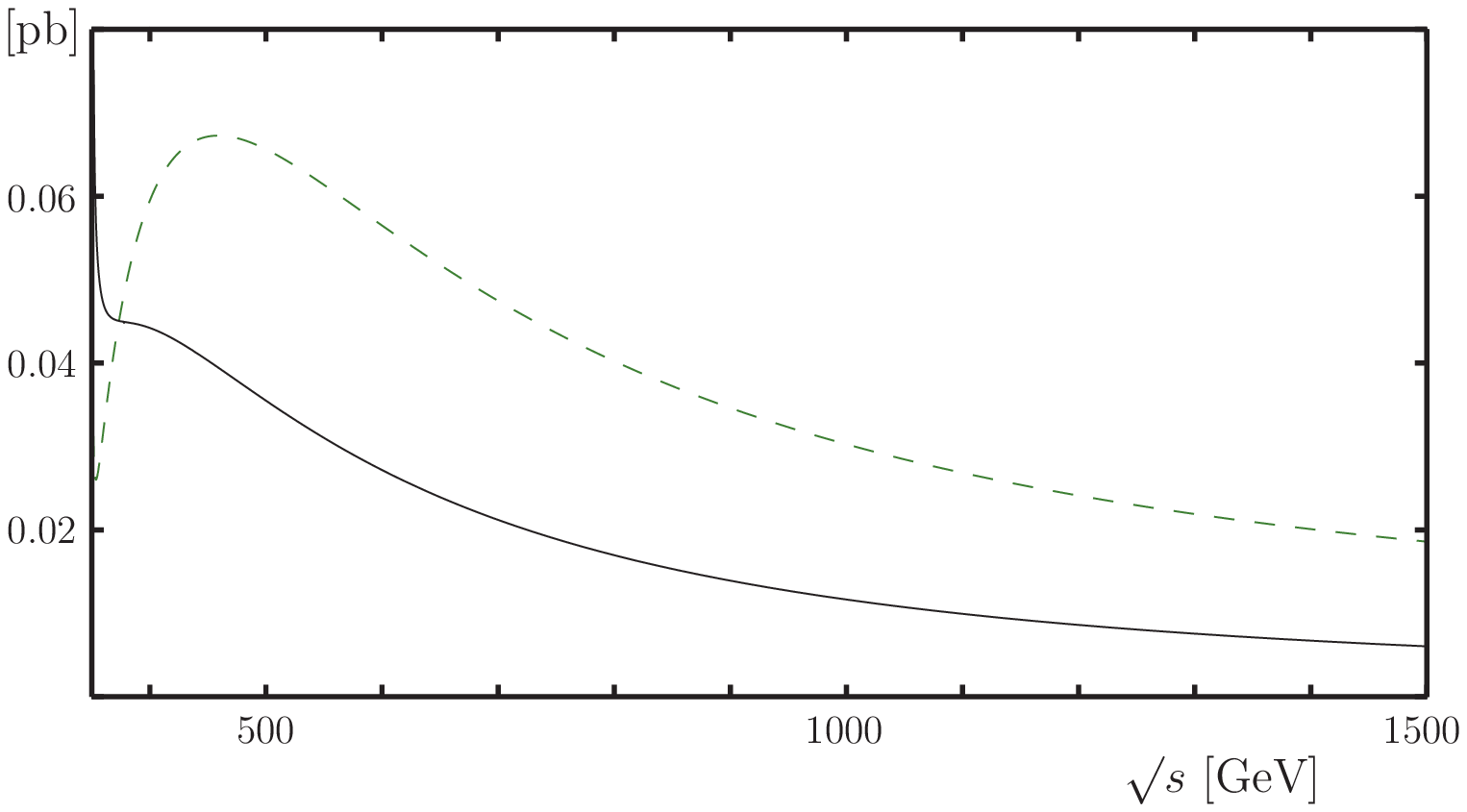}}
\caption{\small Corrections $\sigma_1$ and $\sigma_2$ (dashed and
continuous curves, respectively).\protect\label{Fig3}}
\end{figure}

In Fig.~\ref{Fig1}(a) we present the results of the calculation of
the total cross-section $\sigma(s)$ above the threshold. The
results in percentages with respect to the exact solution are
shown in Fig.~\ref{Fig1}(b). In the latter figure we place also a
result in DPA, where $\sigma_{DPA}(s)$ is determined by the same
formulas as in the case of $\sigma(s)$ but by substituting
$\Phi(s;M^2,M^2)(1+\delta_{c}(s;M^2,M^2))$ for
$\Phi(s;s_1,s_2)(1+\delta_{c}(s;s_1,s_2))$ and $s - M^2 +
\mbox{i}\,\Gamma$ for $\Delta^{-1}(s)$. The similar results for
the hard-scattering cross-section $\hat\sigma(s)$ are presented by
Fig.~\ref{Fig2}(a,b). Let us remember that $\hat\sigma(s)$ is
responsible for the distribution over the invariant mass of the $t
\bar{t}$ system and therefore is of interest, as well
\cite{tt-mass}. In Fig.~\ref{Fig3} we show the results separately
for the NLO and NNLO corrections to $\sigma$. In Table~\ref{tabl}
the results are represented in the numerical form at the
characteristic energies accessible at the planned $e^{+}e^{-}$
colliders. In the last column the numbers in parenthesis represent
the uncertainties in the last digits. (See discussion of their
determination in \cite{ACAT}.) In the other columns the
uncertainties are omitted as they appear in the digits that are
not shown. In the lower positions in the Table the results are
presented in percentages with respect to the exact result in the
model.

\begin{table}[t]
\begin{center}
\caption{\small The results of the calculation of the total
cross-section in pb.}
\begin{tabular} {ccccc}\hline
\hline\noalign{\medskip} \\[-6mm]
 $\quad \sqrt{s}$ (TeV) $\qquad$   & $\qquad \sigma \qquad\;\;$    &
 $\qquad \sigma_{0} \qquad\;\;$        & $\qquad \sigma_{01} \qquad\;\;$ &
 $\qquad \sigma_{012}   \qquad\quad$ \\
\hline\noalign{\medskip} \\[-6mm]
 0.5                               & 0.6724         &
 0.5687          &  0.6344         & 0.6698(7)          \\
                                   & {\small 100\%} &
 {\small 84.6\%} & {\small 94.3\%} & {\small 99.6(1)\%} \\
\hline\noalign{\medskip} \\[-6mm]
 1                                 & 0.2255         &
 0.1821          &  0.2124         & 0.2240(2)          \\
                                   & {\small 100\%} &
 {\small 80.8\%} & {\small 94.2\%} & {\small 99.3(1)\%} \\
\hline\noalign{\medskip} \\[-6mm]
 3                                 & 0.03697        &
 0.02363         &  0.03377        & 0.03653(3)          \\
                                   & {\small 100\%} &
 {\small 63.9\%} & {\small 91.4\%} & {\small 98.8(1)\%}   \\
\hline\noalign{\medskip} \\[-6mm]
 5                                 & 0.02032        &
 0.00904         &  0.01705        & 0.01991(2)          \\
                                   & {\small 100\%} &
 {\small 45.5\%} & {\small 83.9\%} & {\small 98.0(1)\%} \\
\noalign{\smallskip}\hline\hline
\end{tabular}\label{tabl}
\end{center}
\end{table}

The above outcomes exhibit very stable behavior of the NLO and
NNLO approximations in the energy region beginning with
approximately 400~GeV. (In this region simultaneously the right
hierarchy of the corrections is established, $\sigma_0 < \sigma_1
< \sigma_2$.) The accuracy of the NNLO approximation is
established greatly high in this region. In particular, at 400~GeV
$< \sqrt{s}\, <$ 600~GeV it is within $\pm 0.5 \%$. At increasing
energy the accuracy in relative units is slightly decreasing, but
the cross-section is decreasing, too, so that the effective
precision of the description remains approximately the same
(because the ratio of the discrepancy to the quantity
$\sqrt{\sigma}$, which characterizes statistical error, is
approximately constant). In contrast to the above picture, the DPA
exhibits greatly unstable behavior; its discrepancy varies from
+3.0\% to -7.4\% in the energy region 400 GeV $<\sqrt{s}\,<$
1500~GeV. On the whole, such a behavior coincides with the
expected one for DPA, including the order of magnitude of the
discrepancy in the Born approximation for $\Phi$ \cite{LEP2b}.

To conclude this section three important remarks are in order.
First we note that the results expressed in relative units are
almost insensitive to the choice of the test function $\Phi$. In
particular, the turning-on/off of the Coulomb factor has very
small effect. For instance, at $\sqrt{s} = 500$~GeV this leads to
0.7\%-modification of the ratio $\sigma_{012}/\sigma$ and at
$\sqrt{s} = 1500$~GeV does less than 0.1\%. (Although, the
variation of the absolute value of the cross-section is
considerable in both cases: about 35\% and 20\%, respectively.)
The second remark concerns a large value of the correction
$\sigma_2$ in comparison with the discrepancy
$\sigma\!-\!\sigma_{012}$. For example, at $\sqrt{s} = 500$~GeV
they constitute 5.3\% and 0.4\% of $\sigma$, respectively. However
we think that this is an incidental unbalance as $\sigma_2$ gains
its value mainly due to the correction $\alpha^2 \Gamma_2$ to the
width, which in our case exhausts the corrections, and
simultaneously $\alpha^2 \Gamma_2$ is quite large (approximately
2\% of $\Gamma$). If we put everywhere $\alpha^2 \Gamma_2=0$, then
at $\sqrt{s} = 500$~GeV the $\sigma_2$ decreases to 2.1\% with the
discrepancy remaining within the 0.5\%-interval. On the other
hand, if we put $\alpha^2 \Gamma_2=0$ only at calculating the
coefficients $c_n$ without the change of the model itself, then
the $\sigma_2$ becomes almost the same as in the latter case, but
the discrepancy increases to 3.6\%. The third remark concerns the
ill-convergent property of the MPT in the near-threshold region.
With the energy approaching the threshold the accuracy and
stability of MPT rapidly become worse. This manifests itself in
the violation of the hierarchy of the corrections and then in the
blowing up of the corrections. Actually, this behavior was
predicted in \cite{MPT}. To prevent this difficulty another mode
of the MPT near threshold must be applied \cite{MPT} that implies
Taylor expansion of $\sigma(s)$ both in powers of $\alpha$ and in
powers of $s-4M^2$, where $4M^2$ is the threshold. An alternative
method implies a secondary Dyson resummation in the framework of
the MPT approach \cite{EPJC,Proc}.

\section{Discussion}\label{Conclusion}

Although above calculations have been carried out in the framework
of a model, the obtained outcomes expressed in relative units to a
large extent are model-independent. By this we mean that the
outcomes are weakly sensitive to the choice of the test function
determined by the model under consideration. We verified this
property by carrying out calculations with various test functions
and found that the influence of the test function manifests itself
mainly in a factor common for different contributions to the
cross-section. In particular, even very large variation in the
test function that appear at the turning-on/off the Coulomb
factor, in relative units leads to small modifications of the
outcomes.

On this basis we suppose that the loop corrections to the test
function will lead in relative units to small modifications of the
outcomes, too. In particular, our result about the 0.5\%-accuracy
of the NNLO approximation near the maximum of the cross-section,
should remain in force at turning-on the loop corrections.
Moreover, one can further improve the results if applying the MPT
on the background of the loop corrections only, and considering
the Born contribution in the old fashion with the
Dyson-resummation in the unstable-particles propagators --- on
analogy of actual practice of application of DPA
\cite{LEP2a,LEP2b}. In this case the discrepancy in the MPT
description will be diminished by a factor $O(\alpha)$.

Another aspect of the problem of model-dependence of our results
concerns the corrections to the width of unstable particles.
Recall that these corrections determine coefficients $c_n$, which
are crucial for the definition of MPT expansion. We have
considered the case with rather large corrections to the width
(7\% and 2\% in the NLO and NNLO, respectively). At diminishing
these corrections, the MPT corrections to the cross-section should
diminish, too. At least, we have observed this property in the
framework of the model under consideration. On this ground we can
expect the improving of the precision of description at transiting
from the top quarks to EW-only interacting particles, for instance
to the $W$-bosons, because the corrections to the width are lesser
in the latter case.

As regards the application of our results to the description of
realistic processes with the top-quark pair production, we should
stress that our calculations simulate the main contribution to the
cross-section as they cover the double-resonant contributions. So
on the basis of our results we can judge about the precision that
must be achieved in realistic calculations. Fortunately, the
accuracy of the NNLO approximation detected in our analysis, is
satisfactory from the point of view of the ILC requirements.
Really, assuming that at the ILC several hundred thousands of the
$t\bar t$ events is expected, we conclude that the calculation of
the cross-section is needed with a few per mille accuracy. As we
have seen above, this, in general, is ensured by the NNLO in the
MPT.

In summary, we have shown that the MPT stably works at the
energies near the maximum of the cross-section and above at the
description of the total cross-section for the pair production and
decay of fundamental unstable particles. We have found also that
in the mentioned energy region the MPT provides very good
precision within the NNLO. In particular, at the ILC energies in
the case of the top-quark pair production the NNLO approximation
provides 0.5\%-precision of the description. The further increase
of the precision is possible at the proceeding to the NNNLO,
possible on the basis of the results of \cite{MPT}, or at the
proceeding to the compound use of the MPT, when the loop
corrections are treated completely in the framework of the MPT
while the Born contribution to the cross-section is taken into
consideration in the old fashion with the Dyson resummation in the
unstable-particles propagators. On the whole, the MPT method is a
real candidate for carrying out high-precision calculations needed
for ILC.

\end{document}